\documentstyle[12pt,epsf,epsfig]{article}
\def\be{\begin{equation}}
\def\ee{\end{equation}}
\def\ba{\begin{eqnarray}}
\def\ea{\end{eqnarray}}
\def\la{~\mbox{\raisebox{-.6ex}{$\stackrel{<}{\sim}$}}~}
\def\ga{~\mbox{\raisebox{-.6ex}{$\stackrel{>}{\sim}$}}~}
\def\bq{\begin{quote}}
\def\eq{\end{quote}}

 at 10truept

\newcommand{\beq}{\begin{equation}}
\newcommand{\eeq}{\end{equation}}
\newcommand{\beqa}{\begin{eqnarray}}
\newcommand{\eeqa}{\end{eqnarray}}
\newcommand{\mpl}{M_{Pl}}

\def\la{~\mbox{\raisebox{-.6ex}{$\stackrel{<}{\sim}$}}~}
\def\ga{~\mbox{\raisebox{-.6ex}{$\stackrel{>}{\sim}$}}~}

\def\ltap{\ \raise.3ex\hbox{$<$\kern-.75em\lower1ex\hbox{$\sim$}}\ }
\def\gtap{\ \raise.3ex\hbox{$>$\kern-.75em\lower1ex\hbox{$\sim$}}\ }
\def\gl{\ \raise.5ex\hbox{$>$}\kern-.8em\lower.5ex\hbox{$<$}\ }
\def\roughly#1{\raise.3ex\hbox{$#1$\kern-.75em\lower1ex\hbox{$\sim$}}}

\begin{document}

\thispagestyle{empty}
\begin{flushright}
hep-th/0409226\\ September 2004
\end{flushright}
\vspace*{1cm}
\begin{center}
{\Large \bf Moduli Entrapment with Primordial Black Holes}\\
\vspace*{1cm} {\large Nemanja Kaloper$^{\dagger,}$\footnote{\tt
kaloper@physics.ucdavis.edu}, Joachim
Rahmfeld$^{\ddagger,}$\footnote{\tt joachim.rahmfeld@sun.com} and
Lorenzo Sorbo$^{\dagger,}$\footnote{\tt sorbo@physics.ucdavis.edu}}\\
\vspace{.5cm} {\em $^{\dagger}$Department of Physics, University
of California, Davis,
CA 95616}\\
\vspace{.15cm} {\em $^{\ddagger}$Sun Microsystems, Inc., 7777
Gateway Boulevard, Newark, CA 94560}\\
\vspace{.15cm} \vspace{2cm} ABSTRACT
\end{center}
We argue that primordial black holes in the early universe can
provide an efficient resolution of the Brustein-Steinhardt moduli
overshoot problem in string cosmology. When the universe is
created near the Planck scale, all the available states in the
theory are excited by strong interactions and cosmological
particle production. The heavy states are described in the low
energy theory as a gas of electrically and magnetically charged
black holes. This gas of black holes quickly captures the moduli
which appear in the relation between black hole masses and
charges, and slows them down with their {\it vevs} typically close
to the Planck scale. From there, the modulus may slowly roll into
a valley with a positive vacuum energy, where inflation may begin.
The black hole gas will redshift away in the course of cosmic
expansion, as inflation evicts the black holes out of the horizon.

\vfill \setcounter{page}{0} \setcounter{footnote}{0}
\newpage

Inflation \cite{inflation} is at present the best framework for
explaining the origin of our universe. However there still are
technical difficulties with implementing inflation in fundamental
theory. Inflation needs light scalar fields with very flat
potentials and positive vacuum energy, which are hard to obtain
from first principles. Even if such degrees of freedom are
present, there is still the problem of arranging favorable initial
conditions \cite{initial}, which ensure that the vacuum energy
controlled by the light scalars dominates over other energy
sources in the inflationary patch. One approach towards resolving
this problem is the idea of eternal inflation
\cite{steinh,vilenkin,chaotic} (for a recent review see
\cite{alan}), which posits that after inflation starts in some
regions of a huge ``metauniverse" where the environmental
conditions favor it, it will produce many universes as big as
ours.

Recently, there has been progress in the direction of implementing
eternal inflation in string theory. Backgrounds where eternal
inflation can occur have been constructed in flux
compactifications \cite{kklt,kklmmt,race}. These solutions
naturally belong to the landscape of string vacua
\cite{bopo,Susskind,douglas,savas}. In fact, many of the older
cosmological solutions in supergravity limits of string theory
without future singularities \cite{oldsols} could be fitted at the
foothills of the landscape, close to the supersymmetric limits.
The landscape picture provides for many valleys where inflation
can happen, with some of the moduli fields as inflatons. This
would realize the old hope that string moduli may be inflatons
\cite{bingai}. However, this approach may also resurrect the
moduli overshooting problem, discussed by Brustein and Steinhardt
\cite{bruste}. Depending how they start, the moduli fields may
acquire a lot of kinetic energy on the approach towards the
inflationary minima and overshoot them, running off towards the
supersymmetric regions of weak coupling. There the spacetime
decompactifies and new dimensions of space open up. Thus in order
to find a phenomenologically viable cosmology, it is essential to
find a way to gently lower the moduli to where inflation can occur
and the world remains four-dimensional for a long time.

In this framework, the universe may be described as a compact
space, with flat or negatively curved spatial slices
\cite{andreicomp} (see also \cite{coulemartin}). It is created
close to the Planck scale. Three of the spatial dimensions grew
very big during inflation. The modulus which plays the role of the
inflaton could start in the inflationary valley of the potential
like in topological inflation \cite{race,andreicomp}. However in
some regions modulus may have started high up the slope, and
normally it would have overshot the minimum as it rolled down; if
instead there is enough radiation, the modulus could still get
captured in the minimum thanks to the extra cosmological friction
\cite{rasha}. The main ingredient of this mechanism has been
observed in string cosmology earlier in \cite{dilatons,tsey}. In
the backgrounds dominated by conformal matter where $T^\mu{}_\mu =
0$ the source terms arising from the gravitational couplings of
the moduli to matter vanish and moduli are quickly kinetically
damped \cite{wetterich,oldconf}. Their energy density dilutes as
$\rho_h \sim 1/a^6$ for the homogeneous mode or as $\rho_i \sim
1/a^4$ for inhomogeneities \cite{banks}, where $a$ is the
cosmological scale factor. Some aspects of more phenomenologically
motivated, but similar, mechanisms have also been discussed in
\cite{tida,bcc,hsow}.

We will present here a different mechanism for moduli entrapment.
The idea is to use the source term coming from the interactions of
the moduli with very heavy, nonrelativistic, charged states in the
early universe to enhance the attraction of the inflationary
basin. To illustrate the mechanism consider the following analogy
with a marble and a bathtub. If one drops a marble alongside the
steep wall of the bathtub, the marble will have acquired so much
kinetic energy that it will fly over any of the small dimples in
the tub's floor and continue rolling around for a long time.
However, if the tub is first filled with water, which slowly
drains away, then the marble will loose its kinetic energy very
rapidly thanks to the interactions with the water molecules. It
will be captured quickly in a dimple on the bottom near the place
of original impact. Then water will drain away, analogously to the
redshifting of the cosmological matter contents, leaving the
marble, or the modulus potential, in control of the evolution of
the universe. We will show that a gas of heavy states in string
theory, described in the low energy supergravity limit as extremal
black holes, can enforce the same effect on the runaway moduli of
string compactifications. In this way, the gas of black holes,
with high initial energy density, can provide a new mechanism for
overcoming the Brustein-Steinhardt overshoot problem.

In what follows we will work with a single dilaton-like scalar
modulus, assuming that the theory has been compactified to four
dimensions already. We believe that the same mechanism can be
realized in more complicated scenarios. Then in the approach of
\cite{andreicomp}, the universe emerges somewhere on the landscape
with an energy density close to the string scale. We will consider
what happens with the regions where the modulus is not immediately
placed in the inflationary plateau. Instead it starts far from the
minimum along a steep slope and acquires a large kinetic energy.
Because the spatial curvature of the universe is either zero or
negative, it does not immediately collapse, but expands under the
influence of the dominant sources of stress-energy. This separates
the flatness problem from the Planck scale physics, and allows for
the possibility that inflation may begin well below the Planck
scale. At very high energies where the universe is born, strong
interactions, and specifically gravitational particle production
\cite{gpp,zelsta} will excite all the modes in the spectrum of the
theory. This includes the heavy states in the theory, with masses
above the string scale. When their masses exceed the Planck scale,
such states, with some given quantum numbers, are described as BPS
black holes in the supergravity limit. Such states are known to
locally fix the scalar field {\it vevs} on the horizons to values
completely independent of the asymptotic geometry \cite{fixed}.

The masses of these states depend explicitly on the expectation
value of the scalar at infinity. To see this, recall the example
of the BPS black hole solutions in heterotic string theory,
described by the bosonic sector of the effective supergravity
action \cite{gibbons,klopp,dura}
\be S = \int d^4 x \sqrt{g} \, \frac{e^{-\varphi}}{\ell_S{}^2} \,
\Bigl(R + (\partial \varphi)^2 - \frac12 F^2_{\mu\nu} \Bigr) \, .
\label{sugra} \ee
A generic BPS soliton of (\ref{sugra}) is described by an extremal
black hole configuration with mass ${\cal M}$, electric charge
${\cal Q}$ and magnetic charge ${\cal P}$, which are related by
\cite{gibbons,klopp,dura}
\be {\cal M} = \frac12 \, \Bigl( |{\cal Q}| e^{\varphi/2} + |{\cal
P}| e^{-\varphi/2} \Bigr) \, , \label{mass} \ee
where $\varphi$ is the dilaton {\it vev} infinitely far away from
the hole, determining the asymptotic value of the string coupling
$g_S = \exp(\varphi/2)$. This parameter is left completely
undetermined by the local black hole dynamics. In the
supersymmetric limit it is a modulus which can take any value
because it is a flat direction of the theory. The charges ${\cal
Q}$ and ${\cal P}$ are quantized in the units of the string scale
$\sim 1/\ell_S$. Similar situation persists for the case of other
moduli fields. If, for the sake of simplicity, we assume that all
but one of the moduli are stabilized in the compactification to
four dimensions (such as in \cite{kklt,gkp}), the effective 4D
action which describes the light modes of the theory is, in the
Einstein frame, \cite{gibbons}
\be S = \int d^4 x \sqrt{g} \, \Bigl(\frac{R}{2 \kappa^2} -
\frac12 (\partial \phi)^2 - \frac{1}{4} e^{-{\tt g}_2 \kappa \phi}
F^2_{\mu\nu} \Bigr) \, , \label{dila} \ee
where ${\tt g}_2$ is the modulus coupling, and $\kappa =1/M_{Pl}$
is the inverse Planck mass in 4D. A generic BPS soliton of the
theory (\ref{sugra}) is described by an extremal black hole
configuration with mass ${\cal M}$, electric charge ${\cal Q}$ and
magnetic charge ${\cal P}$. In general, the solutions for the
arbitrary charge assignment for ${\cal Q}, {\cal P}$ are not known
exactly for an arbitrary value of ${\tt g}_2$ \cite{gibbons}.
However, it is straightforward to solve (\ref{dila}) for the black
hole solutions with only electric, or only magnetic charge, in the
extremal limit. They are given by zero entropy, zero temperature
configurations\footnote{They are null naked singularities, unlike
the Schwarzschild solutions whose interplay with scalars has been
studied in \cite{lyova}. Hence they will not Hawking-evaporate.
This is why they are identified with heavy states of the theory.}
of the same global structure as those with ${\tt g}_2 = \sqrt{2}$
\cite{gibbons}. Their masses and charges are related according
to\footnote{We have absorbed any irrelevant numerical factors into
the definition of couplings.}
\be {\cal M}_q =  \frac{|{\cal Q}|}{\kappa} e^{{\tt g}_2 \kappa
\phi/2} \, , ~~~~~~~~~~~~ {\cal M}_p = \frac{|{\cal P}|}{\kappa}
e^{-{\tt g}_2 \kappa \phi/2}
 \, , \label{masa} \ee
where again $\phi$ is the scalar {\it vev} infinitely far away
from the hole, determining the asymptotic value of the coupling
constant $g = \exp({{\tt g}_2 \kappa \phi/2})$, and can take any
value in the supersymmetric limit. In what follows we will work
with the models where the Planck scale and the string scale are
close together, so that asymptotically $g \sim {\cal O}(1)$,
because typically the minima of the nonperturbative scalar modulus
potential appear in this regime. Appropriately, as we will see in
what follows, the black hole toy model works as a moduli stopper
most efficiently precisely in this regime.

The scalar minima in the mass function (\ref{mass}) have prompted
a suggestion \cite{renata} that moduli could acquire effective
potentials generated by the virtual black hole loops, which can
stabilize them. This idea is very interesting because it may
incorporate some nonperturbative quantum gravity effects, even
though is difficult to reliably compute them. We still don't know
quantum gravity Feynman diagrams for black hole states. Instead of
following this route, we will focus on a different application of
the minima of (\ref{mass}). We will consider an early universe
environment where there is a large number of black holes with
scalar-dependent masses as in (\ref{masa}). Once there is a net
energy density of them, with both electric and magnetic charges,
their interactions with the scalar will produce a damping effect,
trapping the scalar to {\it vevs} of order of the Planck scale
(after we normalize the scalar field canonically). The electric
charges block the modulus from running off to the strong coupling,
where they are heavy, and the magnetic charges block it from going
to the weak coupling, where they become heavy. Thus as long as the
charges hang around the modulus cannot easily escape to either
side. This is akin to the environmentally induced stabilization
effects of \cite{tida}. In what follows we will only consider the
effects of the black hole gas populated with only electric and
only magnetic charges. The dyonic solutions should also exist, and
contribute as well. However in the regime where we will work,
dyons will be generally heavier than monopoles. Because we assume
a nearly thermal initial population of heavy states, we expect
that the dyon contributions and effects will be exponentially
suppressed relative to the monopole states, and we can safely
ignore them.

How do we describe this ensemble of black holes? It is well known
that the extremal black hole states with like charges can be
arranged in static arrays because the electromagnetic repulsions
can neutralize the gravitational attraction between them. In the
early universe these charges would neither remain static, nor
would they all be of the same sign. Gravity respects gauge
symmetries and so charge conservation would require the net charge
to be zero. In general, the black hole states will be initially
produced by strong (gravitational) interactions near the string
scale \cite{gpp,zelsta} both as ``particles" and ``antiparticles",
with opposite charges, carrying both electric and magnetic
charges. Hence we will approximate their initial abundances  by
the nearly thermal distribution law, which should be a sensible
order of magnitude estimate. Systems of such black holes would be
dynamical, since the forces between black holes would not cancel
exactly, due to the presence of charges with both signs.
Furthermore, outside of the black holes the space would not be in
the vacuum, but in a some, roughly thermally, excited state. It is
clear by CPT invariance (\ref{masa}) and equivalence principle
that both ``particles" and ``antiparticles" contribute equally to
the stress-energy. States from all mass scales will contribute to
the total stress-energy. If they meet and annihilate, their energy
would be released as relativistic particles, which we will ignore
later on. The effect of radiation on cosmological moduli dynamics
has already been considered in \cite{dilatons,tsey,rasha}.

While overall unstable, such arrays of black holes may become
sufficiently separated so that the cosmic expansion prevents their
complete annihilation. In this case, away from the static limit
the leading order electromagnetic interactions can be neglected,
and the leading order gravitational effects from the charges can
be approximated by the stress-energy tensor of the fluid of
massive particles \cite{modulispace}. Thus since they are heavy
states, almost immediately after they are produced, surviving
black holes will fall out of thermal equilibrium. Therefore their
leading contribution to the energy density will come from their
rest masses. With this in mind we can use the dilute gas
approximation for the description of the evolution of these
particles. This should be reasonable below (but close to) the
string scale, and will rapidly become better as the universe
expands. The initial energy density of black holes should be high,
but still below the Planck scale, to guarantee the validity of the
(super)gravity description where we can ignore higher derivative
corrections in the effective action.

We expect that while our approach is an approximation, it should
be a reasonable one in most of the space when the initial black
hole density is not too large. Namely, it is clear that close to
any individual black hole the averaging procedure which we will
embrace below must break down, because of the strong fields close
to a hole \cite{fixed}-\cite{renata}. In that region, the solution
is completely controlled by the local dynamics which fixes the
modulus to a value specified by the charges and renders it
insensitive to its value at infinity \cite{fixed,renata}. However
we are interested in showing how an initial population of black
holes can help resolve the overshoot problem \cite{bruste} of
string cosmology, which is the statement that the asymptotic value
of the scalar is really a zero mode that may run off to the weak
coupling regime. If the black hole density is not too large
initially, so that their individual separation is e.g. $ \ga {\cal
O}(10)$ Planck lengths (and initially $1/H_0 \ga \ell_{Pl}$), then
the value of the modulus in the space surrounding them will
rapidly approach its asymptotic value. Our approximation should be
a reasonable approximation controlling the dynamics of the modulus
in the interstitial space between the holes. Once inflation
starts, this approximation will rapidly get better and better as
time goes on. In a more realistic situation, however, the modulus
will have a distribution of values determined by the individual
black holes. There will be pockets with all kinds of values of
couplings locally fixed by hole charges, and those which can be
later stabilized by different minima of the landscape potential
where inflation can occur will give rise to sibling universes with
different low energy couplings\footnote{We thank Andrei Linde for
interesting discussions of this issue.}.

We can estimate the initial energy density of black holes as
follows. The number density of the black hole states which are
produced is Boltzmann-suppressed, $n_{\cal M} \propto \exp(-{\cal
M}/T_0)$, where $T_0 \sim \sqrt{H_0 M_{Pl}}$ is the temperature of
the universe when it is formed. We will take it to be $T_0 \la
1/\ell_S$, so that we can reliably use the supergravity limit for
the description of cosmology. The total energy density which these
massive states contribute will be \cite{wetterich,dilatons}
\be \rho_{BH} = \sum_{{\cal M}}  n_{\cal M} \, {\cal M} \, ,
\label{density} \ee
where we estimate ${\cal M}$ by (\ref{masa}) and $n_{\cal M}$, as
indicated above, roughly by the Boltzmann distribution when ${\cal
M} \ge 1/\ell_S$.  Thus the main contribution will come from the
lightest black holes. From (\ref{masa}) and $T_0 \simeq 1/\ell_S$
we see that the Boltzmann suppression factor of the initial number
density scales as $n_{\cal M} \propto \exp\Bigl(- |{\cal Q}|
M_{Pl} \ell_S g_{0}  \Bigr)$ and $n_{\cal M} \propto \exp\Bigl(-
|{\cal P}| M_{Pl} \ell_S/g_{0} \Bigr)$. For a fixed initial value
of the coupling constant $g_{0}$, the masses obey ${\cal M} \sim
g_{0} {\cal Q} M_{Pl} \sim {\cal P} M_{Pl} /g_{0} \gg 1/\ell_S$,
and since $M_{Pl} \sim 1/(g_{0} \ell_S)$ they indeed are black
holes, ${\cal M} \ga M_{Pl}$. Thus when $g_{0}$ is of the order of
unity, the black holes can be described as a nonrelativistic
fluid, whose pressure can be neglected, and whose energy density
in an approximately FRW background can be estimated from
(\ref{density}) with the help of a saddle point approximation. In
fact because the mass in (\ref{mass}) is a linear combination of
the terms proportional to $g$ and $1/g$, assuming a roughly
thermal initial distribution $n_0 \sim e^{-{{\cal M} (\phi_0)
\ell_S}} / \ell_S^3$ we can sum up the magnetic and electric
contributions to $\rho_{BH}$ separately. This yields the initial
black hole energy density
\be \rho_{BH~0} = \bar n_0 M_{Pl} \Bigl(  \langle q \rangle g_0  +
 \langle p \rangle g_0^{-1} \Bigr) \, , \label{energdense} \ee
where $\bar n_0$ is the number density of the lightest black holes
in the ensemble, and $\langle q \rangle = \sum_{\cal Q} |{\cal Q}|
\exp(-{|\cal Q}| \mpl \ell_S g_0)/\bar n_0$ and $\langle p
\rangle$ are ensemble-averaged values of ${|{\cal Q}|}, {|{\cal
P}|}$ respectively, obeying roughly $\langle q \rangle \sim
\langle p \rangle \sim {\cal O}(1)$ when initial value of $g$ is
not too far from unity. As the universe evolves undergoing cosmic
expansion, the energy density of the black hole gas will change in
two ways: {\it i)} it will redshift away according to the usual
$\sim 1/a^3$ law for massive particles, since we are ignoring the
black hole interactions (we expect that this is justified for
heavy black holes when they are sufficiently separated
\cite{modulispace}) and {\it ii)} the evolution of the modulus
will change the coupling and so the mass of the black holes, as is
clear from (\ref{masa}). When the evolution is smooth, a good
approximation accounting for these two effects is to represent the
energy density as a function of the scale factor and the modulus
{\it vev} as \cite{wetterich,oldconf}
\be \rho_{BH} = \bar n_0 M_{Pl} \Bigl(\frac{a_0}{a}\Bigr)^3
\Bigl( \langle q \rangle  e^{{\tt g}_2 \kappa \phi/2} + \langle p
\rangle e^{-{\tt g}_2 \kappa \phi/2} \Bigr) \, , \label{energdens}
\ee
It is convenient to define  $\phi_*$ by $\exp({{\tt g}_2 \kappa
\phi_*}) = {\langle p \rangle}/{\langle q \rangle}$. Then the
formula for the energy density of black holes becomes
\be \rho_{BH} = \frac{\rho_0}{a^3} \, \cosh(\frac{{\tt g}_2
\kappa( \phi-\phi_*)}{2}) \, , \label{energydens} \ee
where $\rho_0 = \sqrt{\langle p\rangle\,\langle q \rangle}\, \bar
n_0 a^3_0 M_{Pl}$. Here we allow for the change in $\rho_{BH}$ via
a subsequent adiabatic evolution of $\phi$ away from its initial
value. In the limits $g_{0} \gg 1$ and $g_{0} \ll 1$ the number
densities of black holes with large electric charge and with large
magnetic charge, respectively, are exponentially suppressed. We
will comment on these limits later on.

The energy density of black holes (\ref{energydens}) will appear
as a source in the Friedmann equation governing the evolution of
the scale factor of the universe $a$. In the Einstein frame, this
equation is
\be 3 M^2_P H^2 + 3 \frac{k}{a^2} = \frac{\dot \phi^2}{2} +
V(\phi) + \rho_{BH} + \rho_{rad} \, , \label{friedman} \ee
where we have included for completeness the contribution from all
relativistic particles in the universe $\rho_{rad}$. We stress
that we restrict the spatial curvature to be $0,-1$ only thanks to
the arguments of \cite{andreicomp}, which strongly favor these
values at high densities. Thus the collapse is averted. The
kinetic and potential contributions from the scalar are included
as ${\dot \phi^2}/{2}$ and $V(\phi)$ respectively. Now, to find
how the black hole gas affects the modulus, we need the equation
of motion for its zero mode. A simple way to obtain it is to use
the second order Einstein equation for the scale factor, and
resort to the Bianchi identity to find $\ddot \phi$
\cite{wetterich}. The equation for $\ddot a$ is, at the level of
our approximations where $p_{BH} \simeq 0$ and $p_{rad} =
\rho_{rad}/3$,
\be \frac{\ddot a}{a} = - \frac{1}{6M^2_P} \Bigl(2 \dot \phi^2 -
2V(\phi) + \rho_{BH} + \frac43 \rho_{rad} \Bigr) \, ,
\label{ddota}\ee
and hence taking the first derivative of (\ref{friedman}), using
(\ref{energydens}) and eliminating $\ddot a$ from (\ref{ddota})
yields
\be \ddot \phi + 3H \dot \phi + \frac{\partial V}{\partial \phi} +
\frac{{\tt g}_2 \kappa}{2} \, \rho_{BH}
 \, \tanh(\frac{{\tt g}_2 \kappa(\phi-\phi_*)}{2}) = 0
\, . \label{ddotdil} \ee
This is the master equation which controls cosmological dynamics
of the scalar modulus in this problem.

\begin{figure}[thb]
\centerline{\includegraphics[width=0.4\hsize,angle=-90]{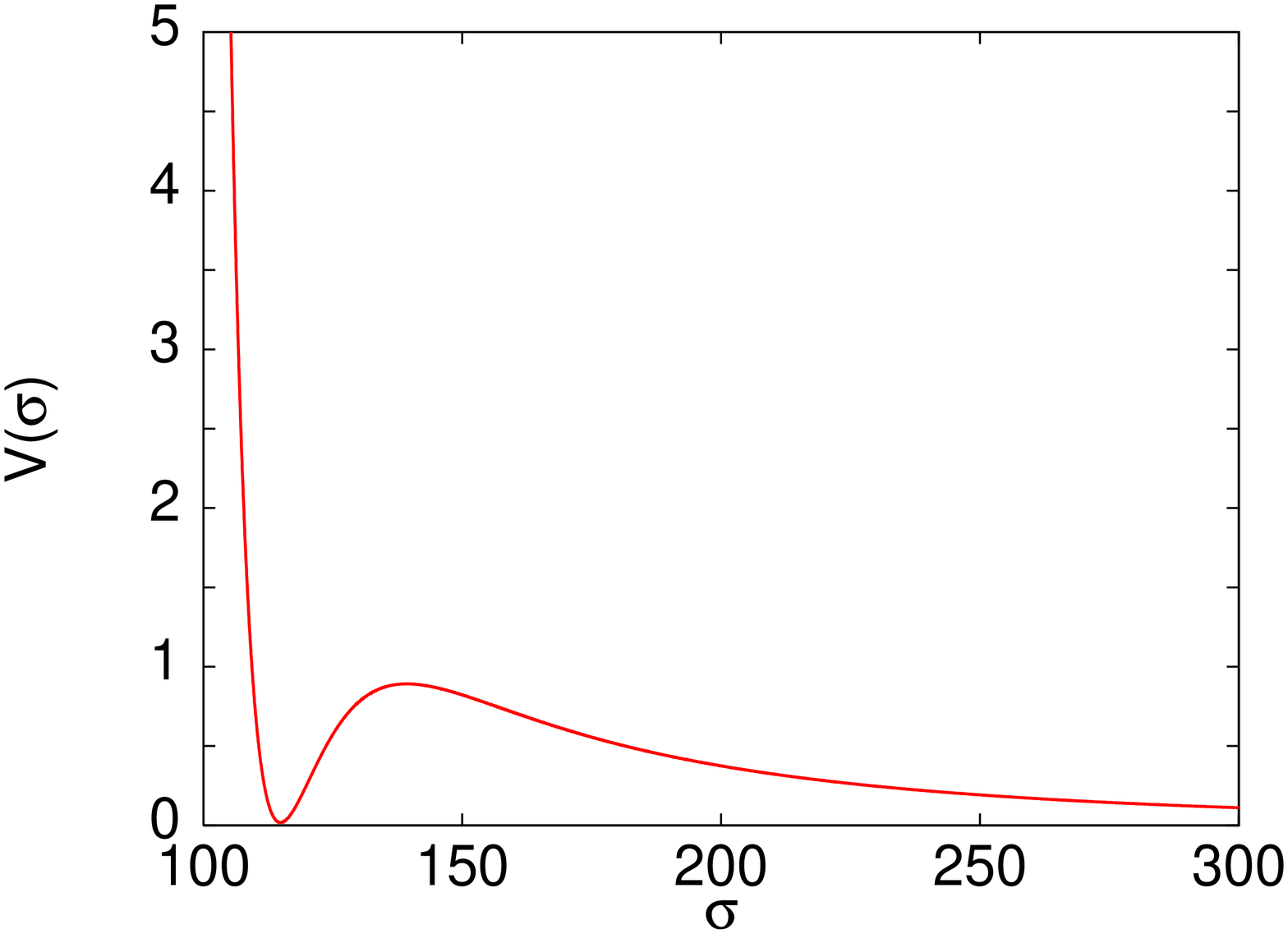}}
\caption{{\small Potential $V\left(\phi\right)$ given by
eq.~(\ref{potential}), in the units of $10^{-15} \mpl^4$. In the
minimum $V \sim \left(10^{-4}\mpl\right)^4 \ne 0$.}
\label{fig:pot}}
\end{figure}

Let us now turn to the dynamical evolution of the universe with a
modulus $\phi$ as governed by equations (\ref{friedman}),
(\ref{ddota}) and (\ref{ddotdil}). First ignore the matter
contributions. A typical potential for the moduli may have some
local minima, but is very steep on one side, and asymptotes as an
exponential function to zero on the other side of the region where
the minima lay. An example is provided by a potential given
in~\cite{kklt}
\be V(\phi) = \frac{a\,A\,{\mathrm e}^{-a\,\sigma}}{2\,\sigma^2}\,
\left[A\,\left(\frac{1}{3}\,a\,\sigma+1\right)\,{\mathrm
e}^{-a\,\sigma} +W_0\right]+\frac{D}{\sigma^3}\,\,,
\label{potential} \ee
where $\sigma=\exp\left(\sqrt{2/3} \, \kappa \, \phi\right)$. When
we consider detailed examples of the scalar evolution below, we
will take $W_0=-10^{-4}$,~$A=1$,~$a=0.1$,~$D=3\times 10^{-9}$ in
units of $\mpl$ following \cite{kklt}, that gives a local minimum
for the potential at $\sim \left(10^{-4}\mpl\right)^4$. We plot
this potential in Fig. (\ref{fig:pot}). The direction of $\phi
\rightarrow \infty$ corresponds to the weak coupling limit, where
the extra dimensions open up and the moduli are the zero modes
from the higher-dimensional graviton multiplet. If the modulus is
trapped in a minimum, the solution will approximate a 4D world for
a long time, controlled by the tunnelling rate through the
barrier, and allow inflationary regime \cite{kklt}.

However, for a generic initial value of the modulus in the strong
coupling, the modulus will fall down the potential and the
evolution will be dominated by its very large initial kinetic
energy. The potential will remain subleading because it is too
steep: its initial role is to just start the phase of kinetic
domination. By the time the modulus reaches a minimum, it will
have acquired too much kinetic energy from rolling down such a
steep potential. As a result the modulus will skip over the
barrier and continue rolling towards the weak coupling. This is
the origin of the overshoot problem \cite{bruste}. If such
compactifications are to avoid special initial condition, they
need a dynamical mechanism which will arrest the modulus before
the flyover. Otherwise, it will be extremely hard to see how to
ever keep the modulus around the minimum and start early
inflation.

The radiation terms $\propto \rho_{rad}$ in (\ref{friedman}),
(\ref{ddota}) can help to slow down the modulus. Their damping
effects have been studied in \cite{rasha} and have been observed
earlier in \cite{dilatons,tsey}. Because of the conformal
symmetry, the stress-energy tensor of radiation vanishes, and so
the radiation does not source the modulus. This is manifest in
(\ref{ddotdil}) where the radiation terms are absent. Then because
radiation scales only as $\rho \sim 1/a^4$ it will overtake the
scalar kinetic energy, which will damp out as $1/a^6$, and the
modulus will stop.

We note that the curvature term $k/a^2$ will work in a similar
vein, but even more efficiently than radiation, if $k=-1$. In that
case, the curvature term will begin to dominate the evolution of
the universe soon after it came into existence, and lead to the
linear expansion $a \sim t$. This will dilute the kinetic term of
the scalar even faster than radiation. Thus if either of these
terms dominates early on, and the scalar does not overshoot the
minimum before it stops to control $H$, it can get caught and
inflation could begin. However, in the case of radiation, since
the black hole density redshifts only as $1/a^3$, it will overtake
radiation within few Hubble times after the beginning, and if
initial $k$ is small or zero, the black hole gas will play the key
role in the evolution of the modulus. In what follows we will
therefore ignore the $\rho_{rad}$ and $k/a^2$ terms, and instead
focus on the black hole contributions $\propto \rho_{BH}$.

To leading order, the effects of the interactions of the modulus
with the black hole gas $\rho_{BH}$ can be described as an
environment-induced mass term, similar to the interactions
discussed in \cite{tida}, and more recently to the chameleon
fields \cite{justin} (see also \cite{gases}). The presence of this
new mass term $\propto \rho_{BH}$ suggests a new method for
capturing the scalar in a potential minimum. The interactions with
the black hole gas pull it towards $\phi_*$ where it awaits for
the contributions from the potential $V$ to overtake the black
hole terms. When $\phi_*$ is divided from the supersymmetric
region by a barrier in $V$, the scalar will eventually settle into
a minimum of $V$ where inflation can take place. It will not
overshoot the barrier since its kinetic energy will be spent
against the black hole mass pumping. However this effect in
general is not always enough to stabilize the scalar. Indeed, if
the gas gets diluted too fast, the interactions will become too
weak to dissipate the scalar kinetic energy. This can be seen
immediately from the comparison with a simple particle mechanics
with a time-dependent mass. For example, if one considers a
pendulum with a mass term $m^2 = \alpha/t^2$, one finds that the
motion is very unstable, and there are modes which move the
pendulum away from the minimum at zero. However, in our case the
black holes do not dilute dangerously fast!

To see this, consider a borough of the landscape where the modulus
$\phi$ measures the size of the 6D space of a warped
compactification of the type IIB theory according to the
prescription of \cite{kklt,gkp}. Then the gauge field in
(\ref{dila}) can arise from reducing a 10D 3-form such that one
index is in the internal space, and so the corresponding choice
for ${\tt g}_2$ is ${\tt g}_2 = -\sqrt{2/3}$. Recall that the weak
coupling is recovered in the limit $\phi \rightarrow \infty$.
Consider (\ref{ddotdil}) with these parameters when the scalar is
close to the effective minimum at $\phi_*$. The small
perturbations $\vartheta = \phi - \phi_*$ obey the linearized
version of (\ref{ddotdil}). After black holes start to dominate in
(\ref{friedman}), the solution behaves as $a = t^{2/3}$ and
$\rho_{BH} = 3M^2_P H^2 = 4 M^2_P/3t^2$. In this limit the
linearized equation becomes
\be t^2 \ddot \vartheta + 2 t \dot \vartheta  + \frac29
{\vartheta} = 0 \, . \label{lineq} \ee
The solutions for $\phi$ then are
\be \phi = \phi_* + \frac{A}{t^{1/3}} + \frac{B}{t^{2/3}} \, ,
\label{linsols} \ee
and so thanks to Hubble damping the evolution is stable under
small perturbations. This is the key ingredient of our mechanism,
which guarantees that the scalar entrapment to $\phi_* = M_{Pl}
\ln( g^2_{0} {\langle p \rangle}/{\langle q \rangle}) $ is gentle
enough to ensure the loss of the scalar kinetic energy.

\begin{figure}[thb]
\centerline{\includegraphics[width=0.35\hsize,angle=-90]{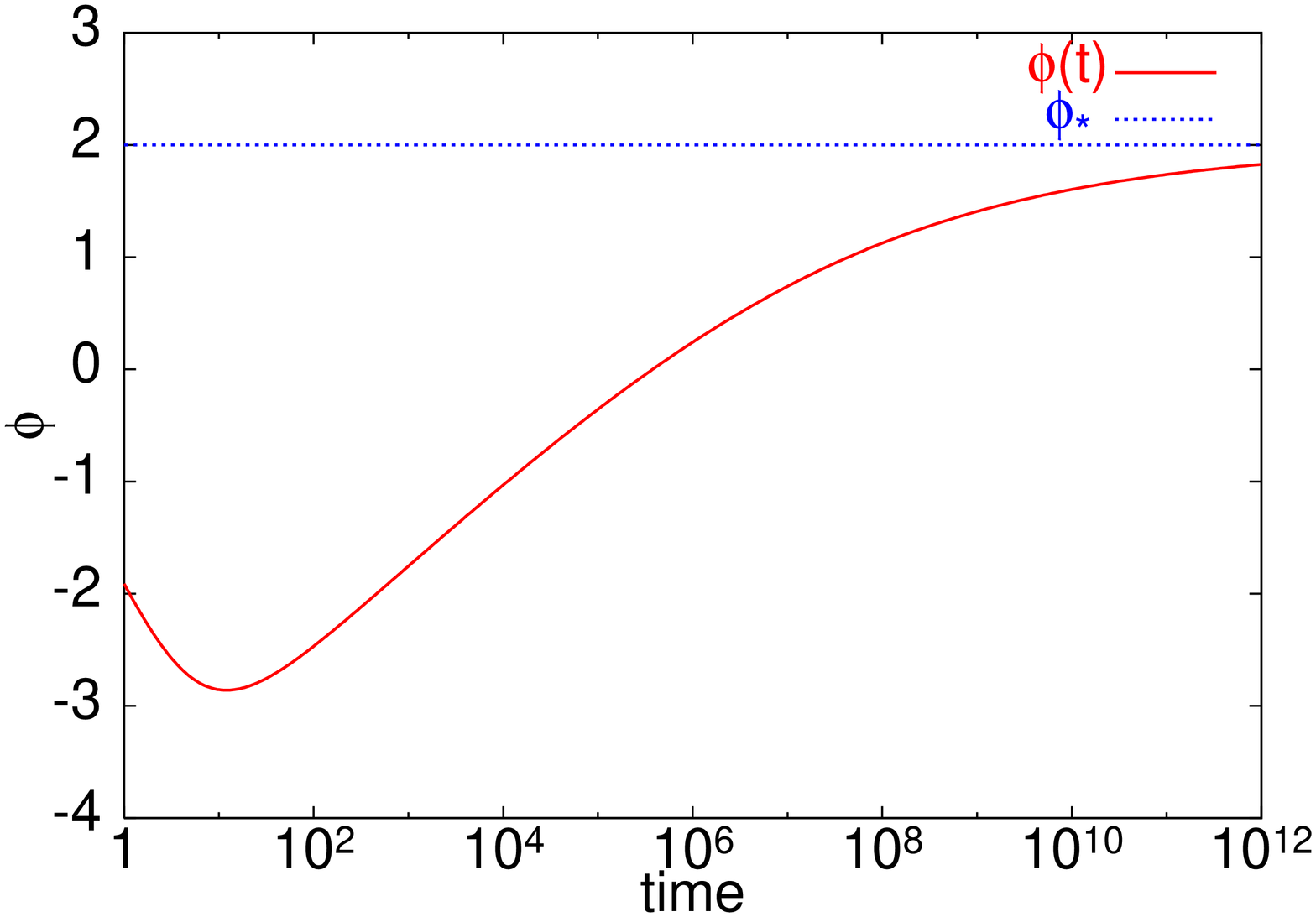}
\includegraphics[width=0.35\hsize,angle=-90]{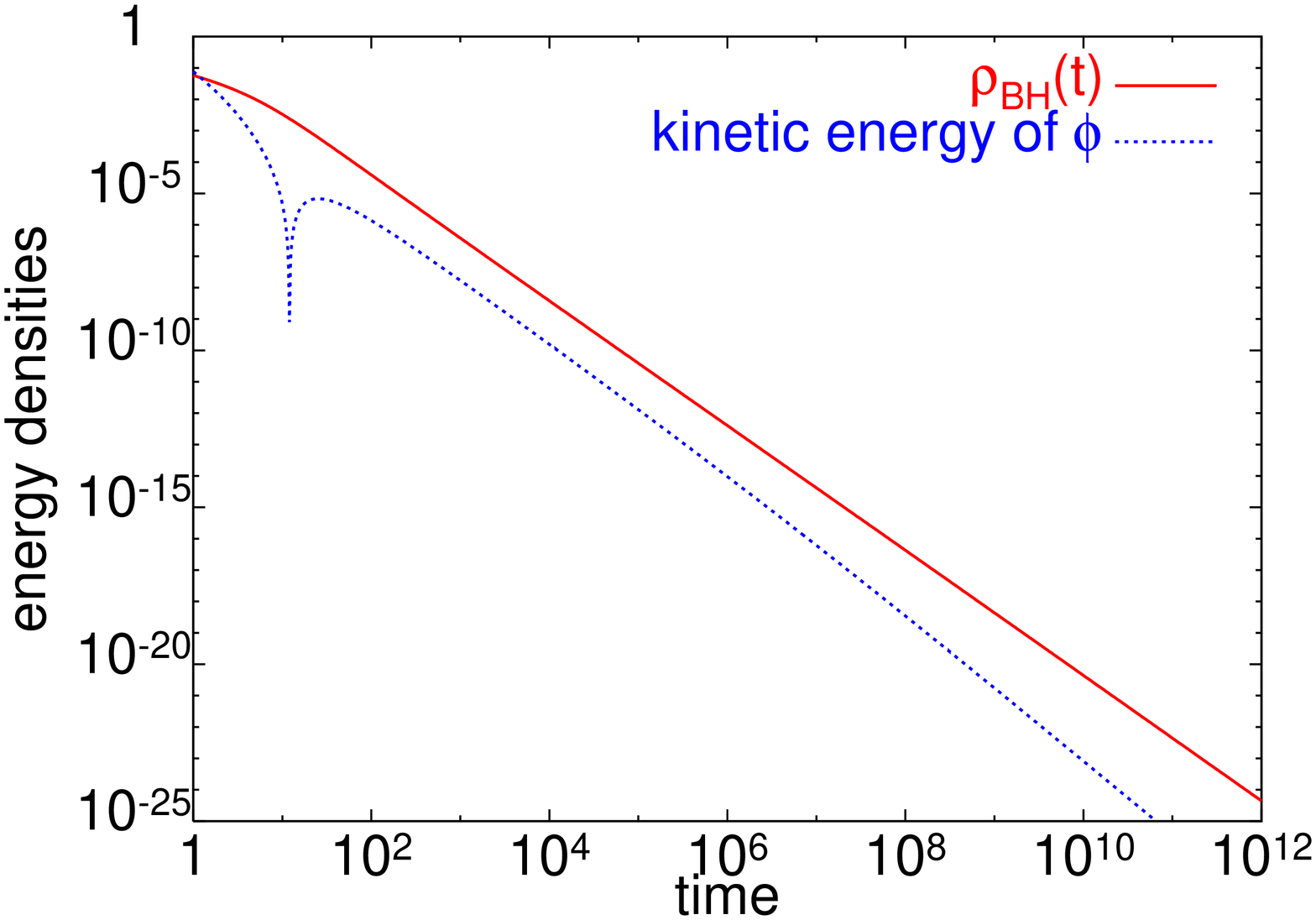}}
\caption{{\small Numerical results for $V=0$. Left: evolution of
the field $\phi$, in the units of $\mpl$, as a function of time.
We depict time in the units of $\mpl/\sqrt{\rho_{BH~0}}$
throughout, and we set $\phi_*=2\,\mpl$. The initial conditions
are $\phi_0=\mpl$,~$\dot{\phi}_0=-\mpl^2$,
$\rho_{BH~0}=10^{-3}\,\mpl^4$ at $t=0$. Right: evolution of
$\rho_{BH}$ and the scalar kinetic energy, in the units of
$\rho_{BH~0}$.} \label{fig:one}}
\end{figure}

We have studied the capture of the modulus by the black hole gas
numerically in order to verify the stability of the entrapment in
the nonlinear regime. In a typical case, the evolution is depicted
in Fig. (\ref{fig:one}). We have taken the initial density of
black holes to be $\sim 10^{-3} \mpl^4$, a number close to, but
safely below the Planck scale, in accordance with our assumptions
that $g_0 \sim {\cal O}(1)$, the validity of the (super)gravity
approximation and that the universe emerged near the Planck scale.
Since the initial effects of the potential are merely to induce a
large kinetic energy of the modulus, early on we can replace it by
picking a large initial value of $\dot \phi \sim M^2_{Pl}$ and set
$V=0$. Further because of the Boltzmann suppression in
(\ref{density}) for massive states, it is natural to take the
initial value of the black hole energy density to be somewhat
lower than $M_{Pl}^4$. In this case the modulus kinetic energy
will start dominating the universe. However it dilutes fast, and
the black hole density starts to contribute quickly. We have
assumed here that the initial value of $g$ is of the order of
unity, so that the location of the attractor for $\phi$ is near
$M_{Pl}$. Clearly, the black holes trap the modulus at a value
$\phi_*$ and the modulus kinetic energy quickly becomes
subdominant.

It is possible to verify this behavior analytically.  An
approximate solution of eqs. (\ref{friedman}), (\ref{ddota}),
(\ref{ddotdil}) initially is given by the FRW cosmology dominated
by a massless scalar, where $\phi(t) \sim
\sqrt{\frac{2}{3}}\,\mpl\, \ln t$, and $a(t) \sim t^{1/3}$. This
ends when the kinetic energy of the scalar redshifts down to the
energy density of the black hole gas, roughly at a time $t_1\simeq
(\mpl/\dot\phi_0) ({\dot\phi_0^2}/{\rho_{BH~0}})^{3/4}$. Since
$\rho_{BH~0}$ is below the Planckian density, $\phi$ will have
rolled several units of Planck mass until black holes begin to
trap it, but not a huge amount. The general dynamics with the
black hole terms is not tractable analytically, however we can
follow the approach towards the attractor at $\phi_*$ as long as
$\phi-\phi_*\gg \mpl$. In this case we can approximate the $\cosh$
and $\sinh$ functions by exponentials, and find a solution
\begin{eqnarray}
\phi\left(t\right)&=&\bar\phi-\frac{2}{5}\,\mpl\,\log\left(1+\mu\,t\right)\,,\nonumber\\
a\left(t\right)&=&a_0\,\left(1+\mu\,t\right)^{8/15}\,,\nonumber\\
{\mathrm
e}^{\bar\phi/\sqrt{6}\,\mpl}&=&\frac{72}{25}\frac{a_0^3\,\mu^2\,\mpl^2}{\rho_0}\,.
\end{eqnarray}
The dimensionful parameter $\mu$ is determined by the scalar speed
$\dot \phi$ at the instant when the black holes begin to take over
from the scalar kinetic energy. This limiting behavior is
consistent with the numerical integration displayed in Fig.
(\ref{fig:one}), describing a slow approach of $\phi$ towards the
attractor $\phi_*$.

\begin{figure}[thb]
\centerline{\includegraphics[width=0.35\hsize,angle=-90]{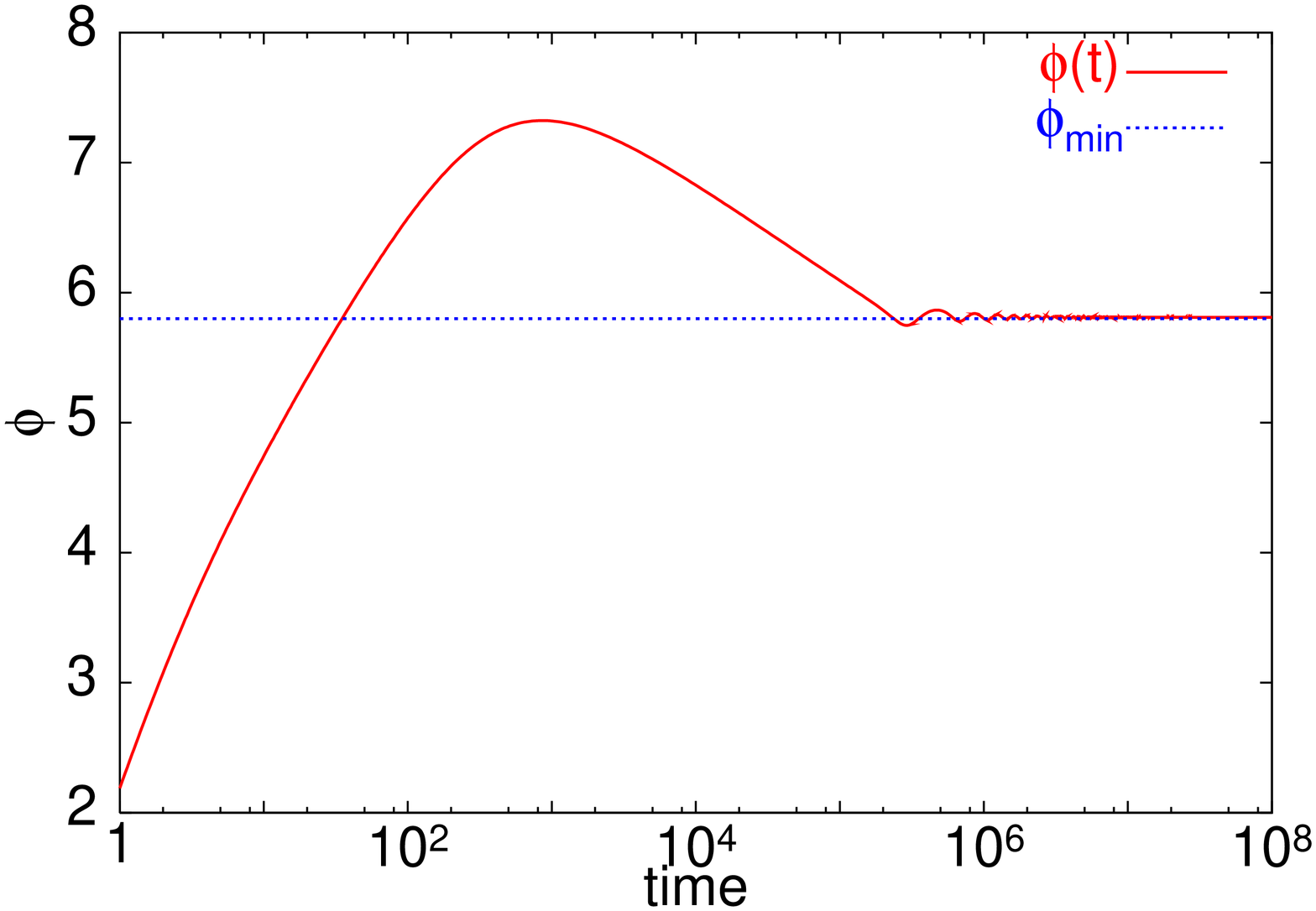}
\includegraphics[width=0.35\hsize,angle=-90]{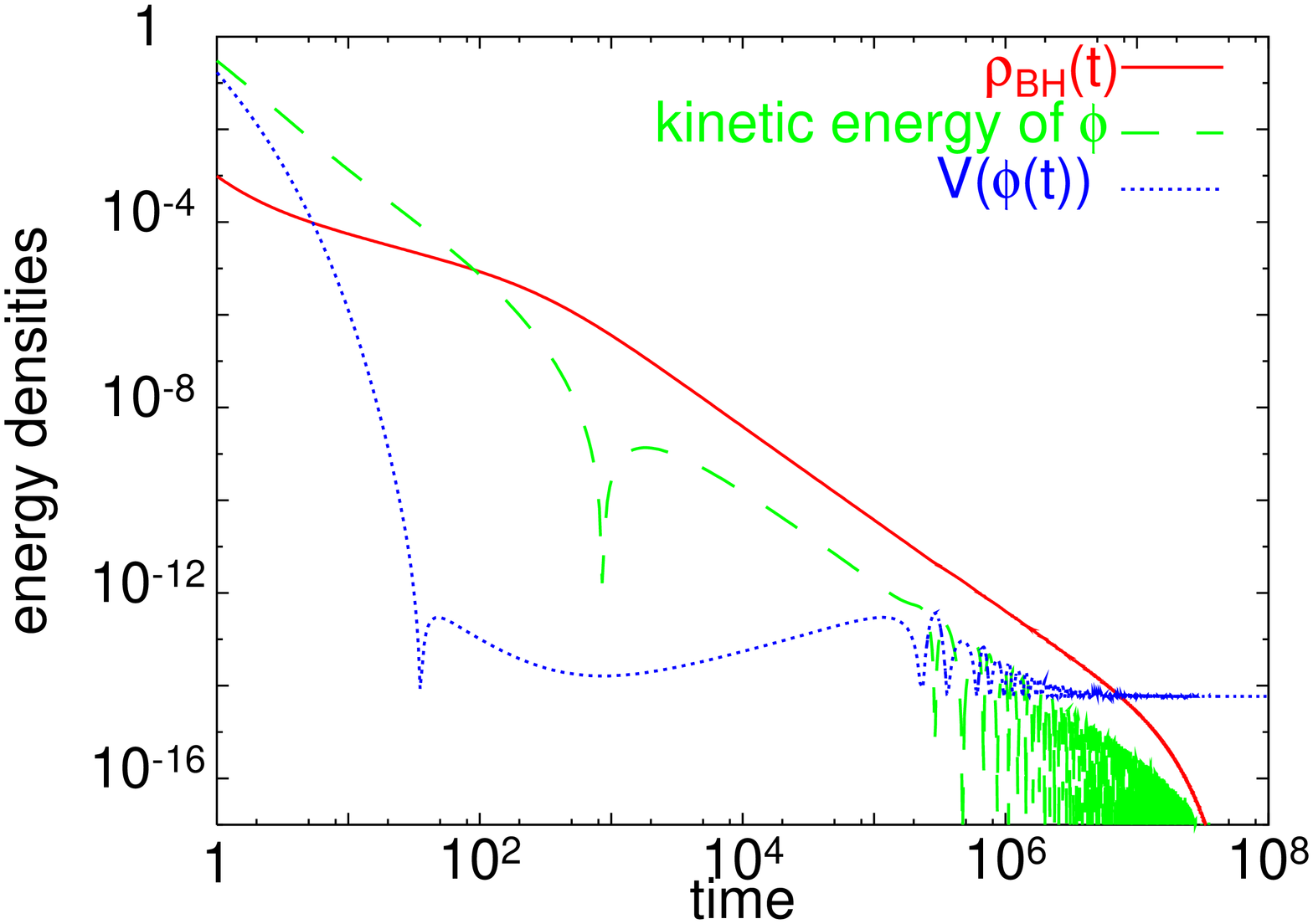}}
\caption{{\small As in Fig. (\ref{fig:one}), but with nonvanishing
potential $V\left(\phi\right)$ given by eq.~(\ref{potential}).
Again $\rho_{BH~0}=10^{-3}\,\mpl^4$. At the time $\sim 10^6$ the
modulus gets trapped in the local minimum of~$V$ located at
$\phi_{\mathrm {min}}\simeq 5.8\,\mpl$. } \label{fig:two}}
\end{figure}

After some time in this regime, the black hole density will have
redshifted to near the scale of the potential
$V\left(\phi\right)$.  If the initial value of $g$ was close to
unity, so that ${\langle p \rangle}/{\langle q \rangle} \simeq
{\cal O}(1)$, the modulus will remain in the region of $\phi \sim
\phi_* \sim {\cal O}(\mpl)$ where the minima of $V(\phi)$ are
located. The modulus will slide down the slope of
$V\left(\phi\right)$ gently, eventually settling down in the
minimum and starting an inflationary era (if the value of $V$ at
the minimum is nonzero, or if the approach to the minimum is along
a very flat slope, where $m_\phi < H$). We plot a typical
evolution of the modulus towards the minimum in Fig.
(\ref{fig:two}). This kind of behavior is generic whenever the
initial black hole density is not too small, and when $\phi_* \sim
{\cal O}(\mpl)$, so that the black hole $\phi_*$ point lies in the
basin of attraction of a minimum of $V$.

\begin{figure}[thb]
\centerline{\includegraphics[width=0.35\hsize,angle=-90]{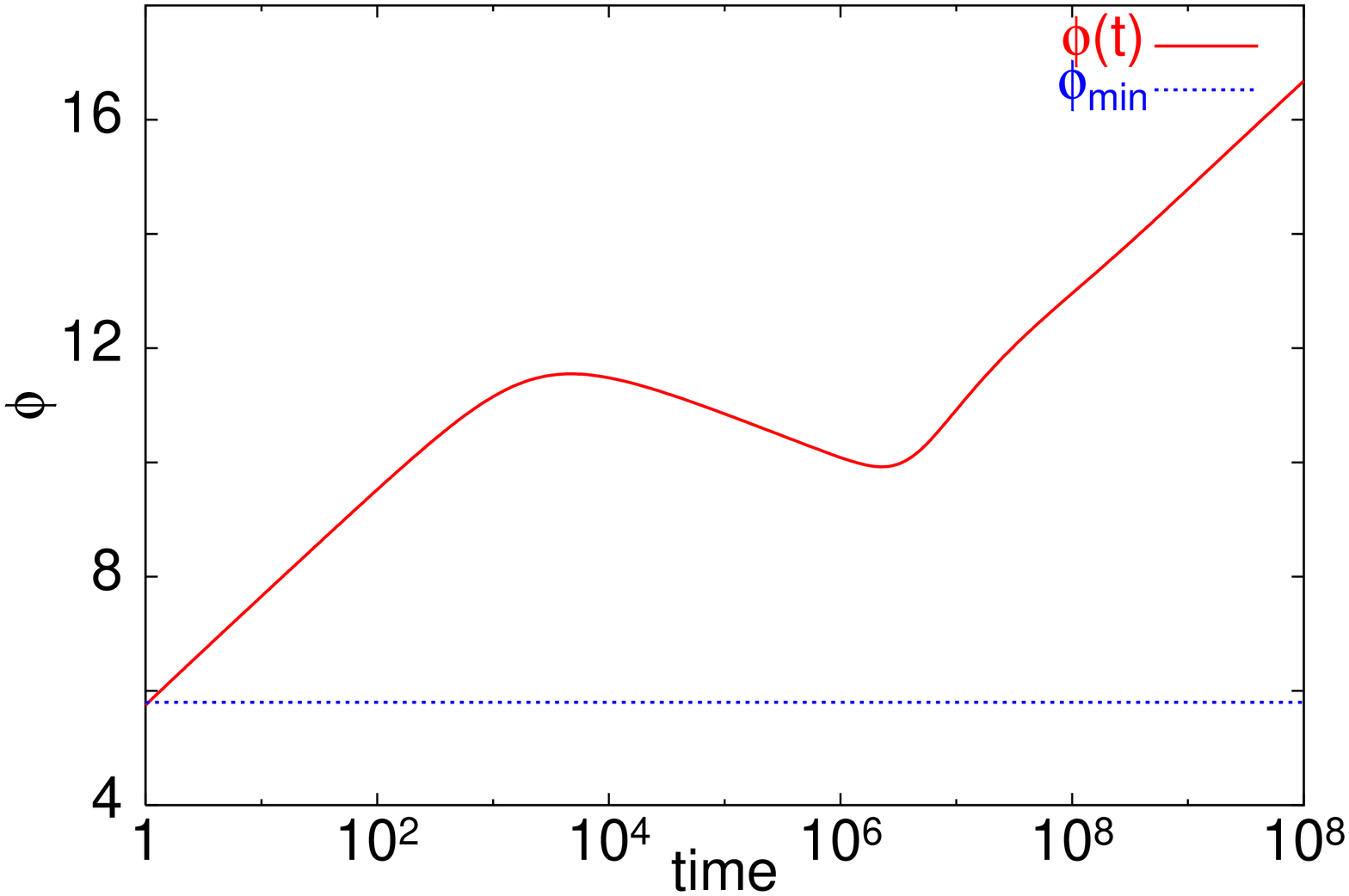}
\includegraphics[width=0.35\hsize,angle=-90]{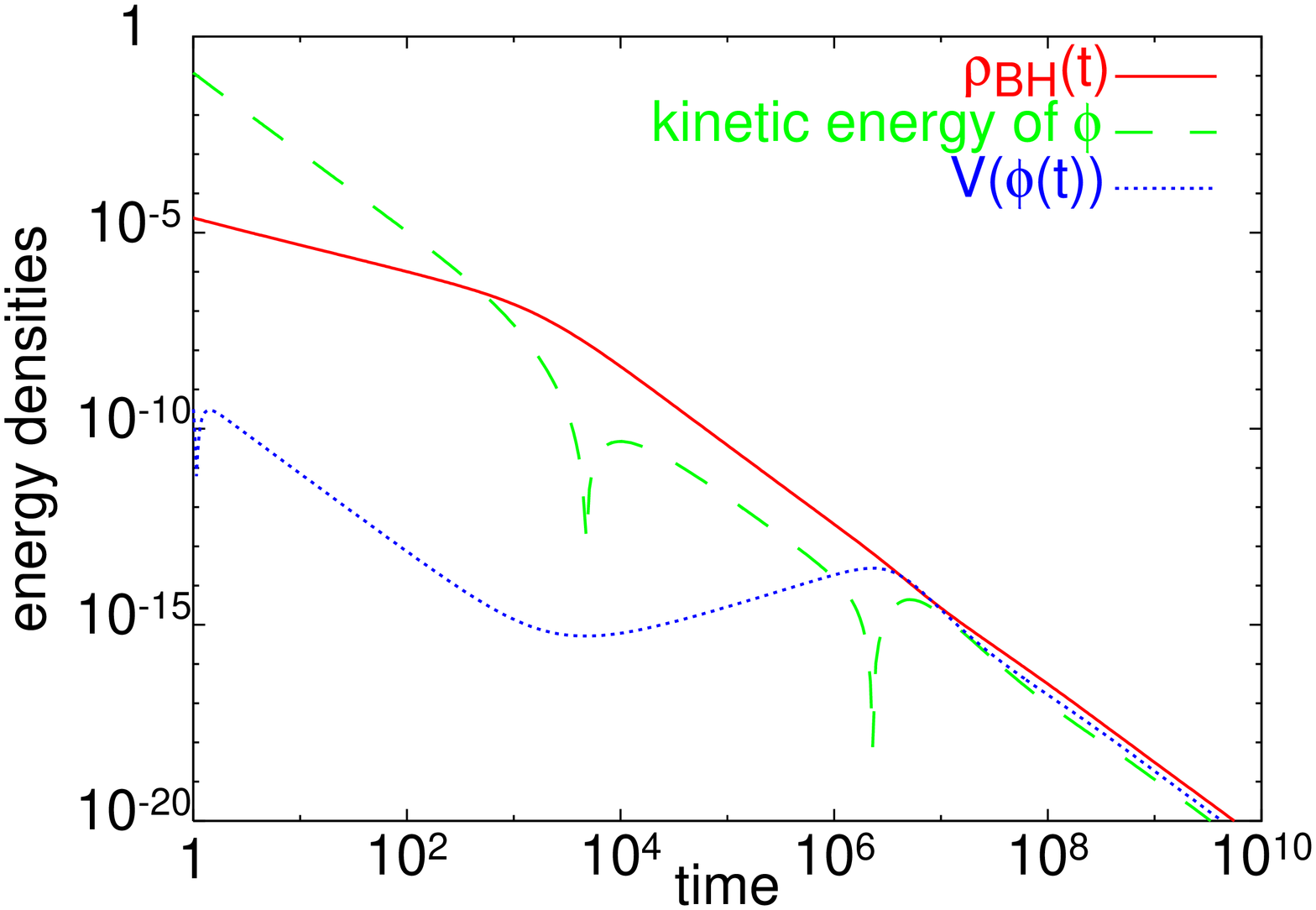}}
\caption{{\small Here we take $\rho_{BH~0}=10^{-6}\,\mpl^4$. Now
the number of black holes is not sufficient to stabilize the
modulus and $\phi$ overshoots the minimum at $\phi_{\mathrm
{min}}$.} \label{fig:three}}
\end{figure}

In the case when the initial value of $\rho_{BH}$ is too small, or
if the attractor $\phi_*$ is far too far in the weak coupling
regime, the modulus will not be trapped in the minimum of the
potential $V\left(\phi\right)$. Either the potential will become
important too early, before the black hole gas manages to slow
down the modulus, or $\phi_*$ will lie outside of the basin of
attraction of the minima of $V$. In each of these cases the
modulus will overshoot the minimum at $\phi_{\mathrm {min}}$ and
will forever continue rolling down the slope $V\left(\phi\right)$
towards the weak coupling. In such situations the
Brustein-Steinhardt problem cannot be avoided and inflation will
not start. The system will eventually converge to a tracking
regime (analogous to the one observed for exponential potentials
in quintessence models). A typical representative of this behavior
is displayed in Fig. (\ref{fig:three}).

However in those cases when the universe is created very close to
the Planck scale, as would typically occur in for example the
approach advocated in \cite{andreicomp}, the initial black hole
density will be significant. This is simply because of
equipartition and equivalence principle. Hence the moduli would be
captured quickly in the black hole attractor $\phi_*$. It still
remains that in the regions of the landscape where the initial
value of the coupling was too small, $\phi_*$ would end up outside
of the basin of attraction and the modulus would still run off to
the weak coupling eventually. However, this behavior is endemic
for the small initial coupling regardless of the subsequent
dynamics. If the coupling had started off too small, the modulus
would never have had a chance of ending in the minimum of
$V(\phi)$ anyway. It would always fall out of the basin of
attraction of the minima of $V$. Thus this initial regime is of
very little interest to start with. One has to make some choices
of the initial conditions, and the whole purpose of the
resolutions of the overshoot problem here and in
\cite{rasha,dilatons,tsey,bcc,hsow} is to demonstrate that the
favorable initial region may be {\it likely}: in this case, it is
a {\it finite} interval of the {\it vevs} of $\phi$. In the
landscape, at very high scales where the influence of the
potential is negligible, the initial value of $\phi$ may be
anything, and so there will be many regions where $\phi \sim {\cal
O}(\mpl)$. In these regions our mechanism will work in a natural
way, and they will inflate. The {\it a posteriori} probability to
be in such regions will then be large because they will become
exponentially big after inflation.

Once inflation starts, the gas of black holes will dilute
exponentially quickly. By the time inflation is over, assuming it
produces about 65 e-folds or so, the number density of the black
holes will drop by a factor of about $10^{90}$. If inflation
happens near the GUT scale, say $V^{1/4} \sim 10^{15} {\rm GeV}$,
the final number density of black holes at the end of inflation
will be
\be n_{BH~final} \le 10^{-90} \frac{V}{\mpl} \sim 10^{-22} {\rm
eV}^3 \,. \label{findens} \ee
After inflation, the universe will expand roughly for another
factor of $10^{90}$ until it reaches the present epoch. The number
density of these primordial black holes today will therefore be no
more than about $n_{BH~final}({\rm today}) \la 10^{-112} {\rm
eV}^3$, leaving us with at most one such black hole per more than
$10^{13}$ present horizon volumes. In other words, these
primordial black holes will become just as efficiently diluted as
the primordial monopoles in the original formulation of inflation
\cite{inflation}. They will quietly go away after completing their
task.

To summarize, in this paper we have proposed a new method for
resolving the Brustein-Steinhardt moduli overshoot problem
\cite{bruste} in string cosmology. We find that a gas of
primordial black holes in the early universe, where the mass of
black holes depends on the modulus, can provide a transient
attractor for the modulus. When such black holes are produced they
will trap the modulus temporarily, and keep it within the basin of
attraction of the minima of the nonperturbative modulus potential
$V$. As time goes on, the black hole density will redshift away,
placing the modulus gently on the potential slope. Following this
the modulus will slowly settle into the minimum, where it can
drive inflation. A key ingredient of our mechanism is that the
universe should be created near the Planck scale, so that the
heavy states are produced initially with non-negligible number
density. Such an approach has been proposed recently in
\cite{andreicomp}, and our mechanism may be a natural ingredient
for helping the modulus stop in the inflationary valleys in the
landscape. After inflation has begun, the density of black holes
redshifts away to exponentially small numbers, just like the
density of heavy monopoles in the early models of inflation. These
black holes are pushed outside of the current horizon. We stress
that our mechanism relies mainly on the presence of heavy states
in the theory whose mass depends on the modulus. It may be
possible to realize a similar mechanism in other regimes, using
different massive states. It would be interesting to consider such
mechanisms in other cosmological models, where for example the
initial density of massive states could be small, but the minima
of the potential lie at very large {\it vevs}, such as in the
theories with low scale unification.

\vskip0.5cm

{\bf \noindent Acknowledgements}

\smallskip

We thank R. Brustein, S. Dimopoulos, R. Kallosh, L. Kofman, A.
Linde and S. Thomas for useful discussions. NK and LS thank the
Aspen Center for Physics for a kind hospitality during the course
of this work. The work of NK and LS was supported in part by the
DOE Grant DE-FG03-91ER40674, in part by the NSF Grant PHY-0332258
and in part by a Research Innovation Award from the Research
Corporation.


\end{document}